\documentclass[paper]{revtex4}

\usepackage{graphicx}
\usepackage{epsfig}
\usepackage{epstopdf} 
\usepackage{color}
\usepackage{xcolor}
\usepackage{amssymb,amsfonts,amsmath}
\begin{document}

\title{Needle-free injection into skin and soft matter with highly focused microjets}

%\author{Yoshiyuki Tagawa\affil{1}{Physics of Fluids Group, MESA Institute and Faculty of Science and Technology, University of Twente, P.O. Box 217, 7500 AE Enschede, Netherlands.} \affil{3}{Email addresses for correspondence: y.tagawa@utwente.nl, c.sun@utwente.nl, d.lohse@utwente.nl},
%Nikolai Oudalov\affil{1}{},
%%Claas Willem Visser (???)\affil{1}{},
%A. El Ghalbzouri\affil{2}{Department of Dermatology, Leiden University Medical Center, Leiden, The Netherlands}
%Chao Sun\affil{1}{}\affil{3}{},
%\and
%Detlef Lohse\affil{1}{}\affil{3}{}}

\author{Yoshiyuki Tagawa$^{1,*}$}
\author{Nikolai Oudalov$^1$}
\author{A. El Ghalbzouri$^2$}
\author{Chao Sun$^{1,*}$}
\author{Detlef Lohse$^{1,*}$}
\affiliation{$^1$Physics of Fluids Group, MESA Institute and Faculty of Science and Technology, University of Twente, P.O. Box 217, 7500 AE Enschede, Netherlands.\\
$^2$Department of Dermatology, Leiden University Medical Center, Leiden, The Netherlands\\
$^*$Email addresses for correspondence: y.tagawa@utwente.nl, c.sun@utwente.nl, d.lohse@utwente.nl,
} 

%\contributor{Submitted to Proceedings of the National Academy of Sciences
%of the United States of America}

%\begin{article}
%\begin{article}
\begin{abstract}

The development of needle-free drug injection systems is of great importance to global healthcare.
However, in spite of its great potential and research history over many decades, these systems are not commonly used.
One of the main problems is that existing methods use diffusive jets, which result in scattered penetration and severe deceleration of the jets, causing frequent pain and insufficient penetration. Another longstanding challenge is the development of accurate small volume injections.
In this paper we employ a novel method of needle-free drug injection, using highly-focused high speed microjets, which aims to solve these challenges.
We experimentally demonstrate that these unique jets are able to penetrate human skin: the focused nature of these microjets creates an injection spot smaller than a mosquito's proboscis and guarantees a high percentage of the liquid being injected. The liquid substances can be delivered to a much larger depth than conventional methods, and create a well-controlled dispersion pattern. Thanks to the excellent controllability of the microjet, small volume injections become feasible. 
Furthermore, the penetration dynamics is studied through experiments performed on gelatin mixtures (human soft tissue equivalent) and human skin, agreeing well with a viscous stress model which we develop. 
This model predicts the depth of the penetration into both human skin and soft tissue.
The results presented here take needle-free injections a step closer to widespread use.
\end{abstract}


\maketitle

\section{Introduction}
The development of needle-free drug injection systems is an essential part of the global fight against the spread of disease \cite{Kane1999, Varmus2003, Hauri2004}. Contamination, needle-stick injuries \cite{Kermode2004}, painful injections, and needle phobia \cite{Nir2003} are issues related to traditional syringe injections with needles that demand attention. Needle-free injections systems offer the prospect of resolving these problems \cite{Mitragotri2005}. Previous studies have explored the possibilities of needle-free injections, but important limitations still need to be addressed \cite{Arora2007, Menezes2009, Stachowiak2009, Mitragotri2006, Baxter2005, Baxter2004}. 

The main issue that is limiting applicability is the shape of the jets produced by the current systems. 
These devices create diffusive jets, leading to a large dispersion pattern and unreliable penetration. 
This in turn creates problems for patients, in the form of frequent bruising and pain \cite{Mitragotri2006}. 
Another problem of conventional methods is that due to the small size of the nozzle diameter, it can easily get clogged, causing disruptions to controllability.

Very recently, we have managed to generate thin, focused microjets with velocities of up to 850 m/s by the rapid vaporization of a small mass of liquid in an open liquid-filled capillary as reported in \cite{Tagawa2012}.
Our novel method of jet creation addresses the issues mentioned above. 
Ultra-high velocities (more than 200 m/s) combined with a highly-focused geometry enable one-shot penetration to the desired area and good controllability. The forces exerted on the liquid that is delivered in this way cause minor damage to the medicine that is contained in it \cite{Hogan2006}.
Due to the fine scale of the jet tip (30 $\mu$m) combined with the high velocities, we can easily adjust the penetration depth according to the requirements. This makes drug delivery efficient and as painless as possible.

In this article we study the penetration dynamics of these highly focused microjets into gelatin mixtures and artificially grown human skin using high-speed imaging.
We investigate the penetration depth as a function of the jet velocity and the capillary tube diameter. 
The understanding of these dynamics will provide essential insight for the development of needle-free injection devices.

\section{Results and Discussion}

\subsection{Injection into gelatin}

\begin{figure*}[ht]
\centerline{\includegraphics[width=.75\textwidth]{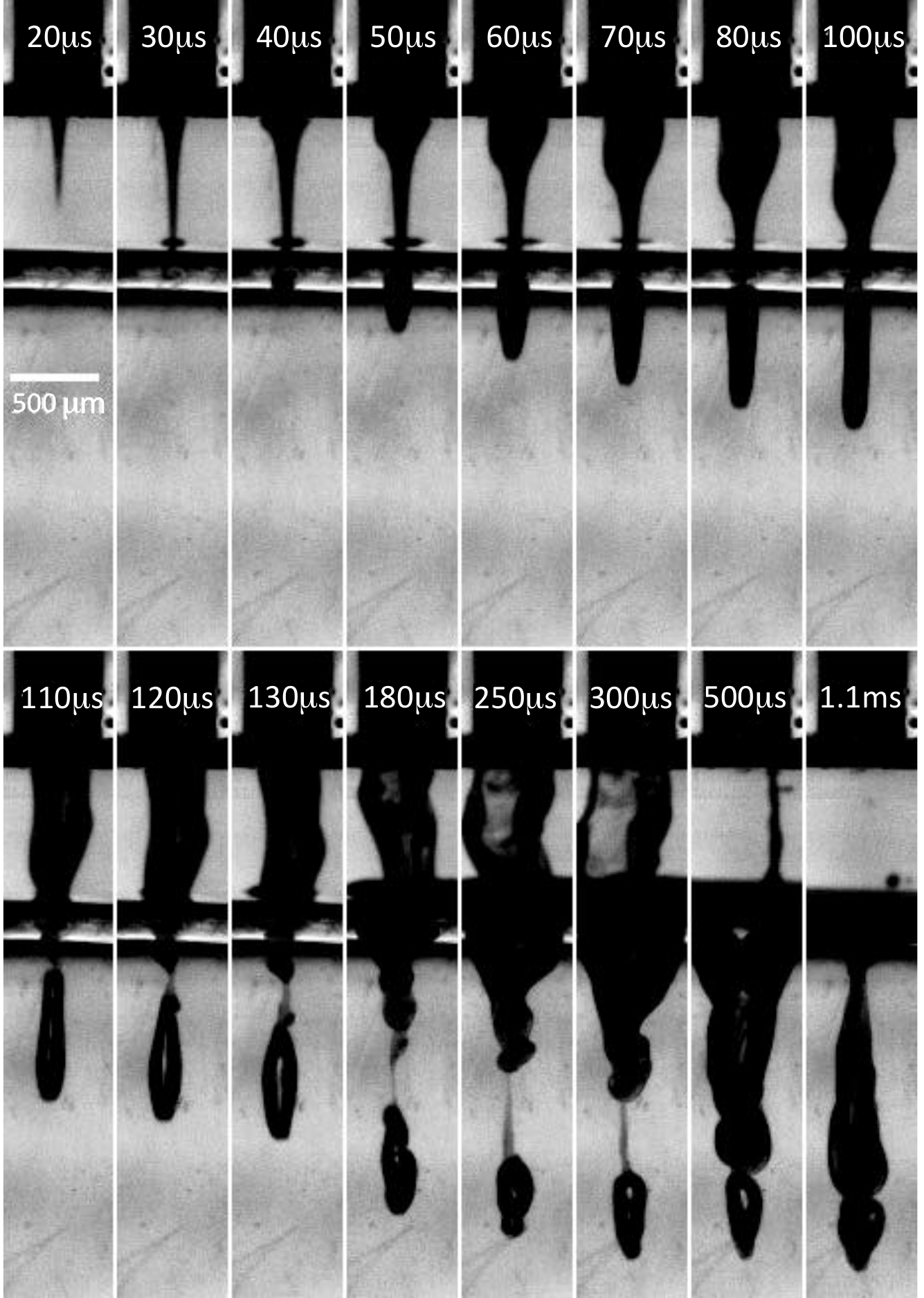}}
\caption{Snapshots of the jet penetration into gelatin. The laser is shot at 0 $\mu$s and the subsequent images show the jet injection process at the designated times. The jet is created in a 500$\mu$m tube.
\label{SAX}}
\end{figure*}

In order to study the penetration of these microjets, we used gelatin 5 wt\% as a model material. This percentage simulates the properties of soft tissue in the human body \cite{Menezes2009}.
Figure~\ref{SAX} provides the first observation of the temporal evolution of the jet penetration.
This visualization is of utmost importance for studying the interaction of the microjet and the human body.

As illustrated in figure~\ref{SAX}, the sharp tip of the microjet reaches the material first.
The diameter of this tip creates an injection spot $\sim$ 30 $\mu$m, smaller than a mosquito's proboscis.
This thin part of the jet starts digging a hole into the material.
Thanks to the highly focused geometry, there is no splashing around the penetration spot, which is crucial for medical applications. This is clearly indicated in figure~\ref{SAX} for the snapshots covering 30 - 100 $\mu$s.
The snapshot at 100 $\mu$s shows a well-controlled dispersion pattern. 
The width of the hole remains as small as the jet diameter.
This is in sharp contrast to the existing methods using diffusive jets, which result in scattered penetration.
The low-speed thick part of the jet utilizes the entry point created by the thin jet and is efficiently deposited into the material.
%Owing to the velocity difference of thin part and thick part, a pinch-off in the gelatin is observed at 110 $\mu$s.
The penetration of the tip stops at about 300 $\mu$s while the thick jet part continues to make its way to the deepest part of the hole.
At the final snapshot (1.1 ms) almost all volume of the jet released by the capillary tube is deposited into the soft material.
The ultrafast jet tip guarantees a high percentage of the liquid being injected as illustrated in figure~\ref{SAX}.
The entire process is finished after 1.1 ms. 

\begin{figure}[h!]
\centerline{\includegraphics[width=0.4\textwidth]{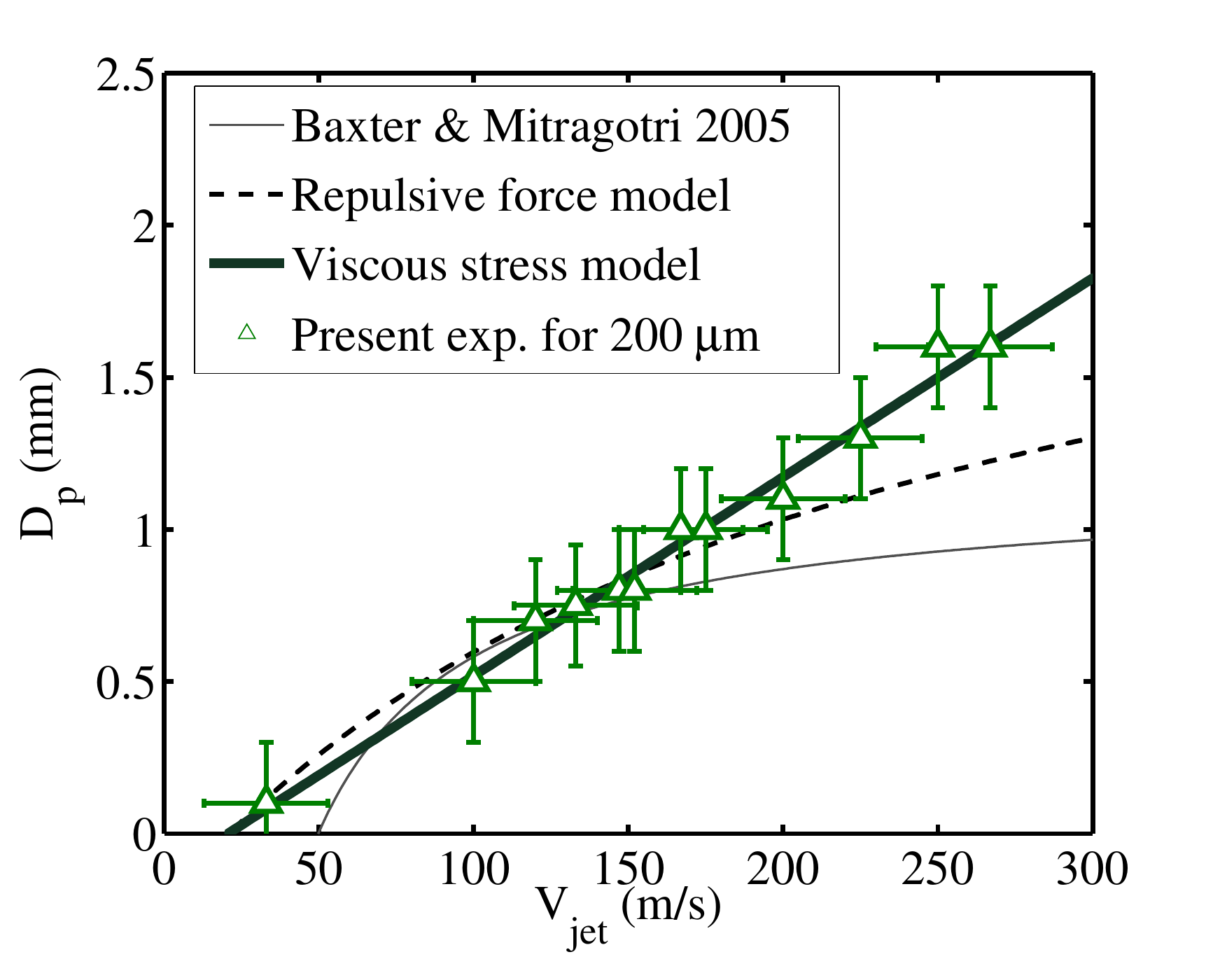}}
\caption{\label{fig:models} Injection depth as a function of the jet velocity. 
Green triangles show the experimental results for the 200 $\mu$m tube.
The depth increases with the jet velocity and no saturation tendency is observed. 
For comparison, the gray thin line shows the Baxter model, the blue dashed line is the repulsive force model, and the black thick line is the viscous stress model with same offset. 
The experimental results agree well with this latter model. %The offset or critical velocity $v_c$ = 21 m/s
}
\end{figure}

Figure~\ref{fig:models} shows the penetration depth of the microjet generated in a 200 $\mu$m tube into gelatin 5 wt\% as a function of the jet velocity.
This penetration is created by a single shot.
The depth linearly increases with the jet velocity, covering depths from several hundred microns at low jet speed to $\sim$1.5 millimeter when the jet velocity approaches $\sim$250 m/s. 
This highlights the versatility of this method, making it adjustable to different skin-properties (e.g. children/adults, different skin types) and to a broad range of medical applications (e.g. insulin injection \cite{Weller1966, Bremseth2004}, vaccinations \cite{Weniger2003, Giudice2005, Kendall2010}, or medical tattoos).

\begin{figure}[h!]
	\centerline{\includegraphics[width=0.32\textwidth]{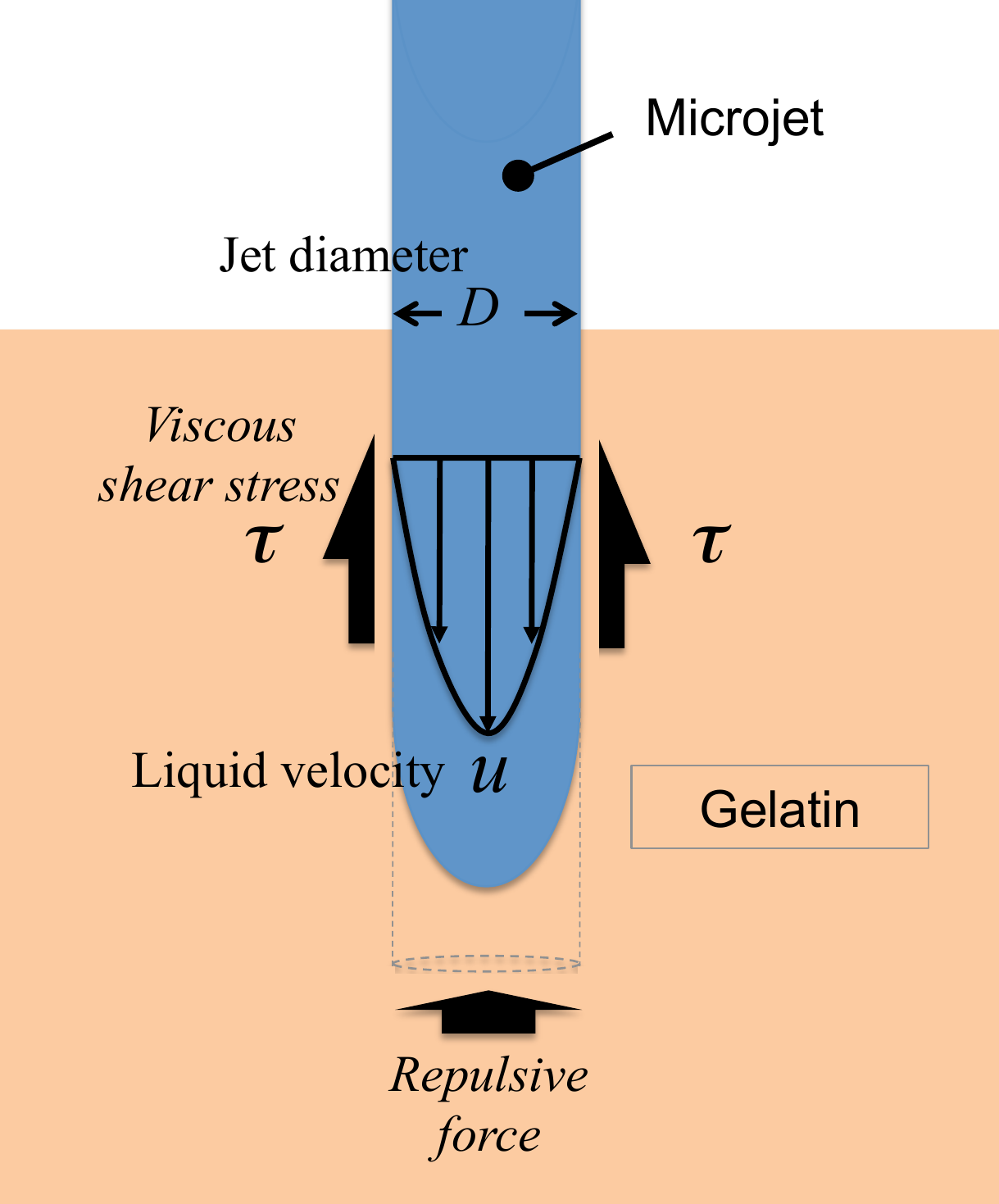}}
	\caption{The schematic sketch of the forces acting on the jet. The jet shown by the blue color region is penetrating into the gelatin. The viscous shear stress $\tau$ acts at the interface between liquid and gelatin due to the shear flow inside the jet. The repulsive force acts vertically on the projection area of the jet shown by the dashed closed line.}
	\label{fig:ModelSketch}
\end{figure}

To get a quantitative understanding, we compare the present results with various models.
A model proposed by Baxter \textit{et al.} \cite{Baxter2005} is fitted to this experimental data and presented in the figure.
The agreement in the low velocity region is fair.
However, this model shows a saturation of the penetration depth for velocities above 200 m/s. In our experiments we did not observe this trend as can be seen in figure~\ref{fig:models}.
%It does not agree with the experiments, i.e. the penetration depth does not show asymptotical behavior while the model shows.
The difference is likely due to the shape of the jet created using our novel method.
The jet shape created by conventional methods using syringe-piston system (e.g. \cite{Arora2007, Menezes2009,Baxter2005}) is diffusive.
This shape leads to severe deceleration with jet travel distance (especially for high velocities), which is considered in the model by Baxter \textit{et al.} \cite{Baxter2005}, resulting in shallow penetration.
On the other hand, the highly focused jets in our experiments do not experience this significant deceleration.

To address this discrepancy we consider the relation between initial impact velocity of the jet and the drag force.
%In general the drag force on an object in fluid dynamical system is described as:
%\begin{equation}
%F_D = \frac{1}{2}C_D \rho  v^2 A,
%\end{equation}
%where $C_D$ is the drag coefficient, $\rho$ is the density of the fluid, $v$ is the speed of the object relative to the fluid, and $A$ is the reference area, often defined as the area of the orthographic projection of the object.
We observe that the gelatin does not show much deformation and the jet penetrates into gelatin with cylindrical shape (see the snapshots at 100 $\mu$s in figure~\ref{SAX}).
We model this phenomena as a cylindrical microjet, normal to the gelatin surface.
It creates a cylindrical `crack' inside the gelatin, which keeps the same circular projection area independent of the depth.
Figure~\ref{fig:ModelSketch} shows the schematic sketch of this model.
The drag forces on the jet are the viscous shear stress at the jet-gelatin interface and the repulsive force on the area of the cross-section of the jet.
%For the present case, two basic models can be considered.
%The dilute concentration of the colloid solution could be regarded as highly viscous liquid. % {\bf (CITATION)}.
%Owing to the dilute gelatin concentration (5 wt\%), 
We consider two basic force models, representing these two different cases: A viscous stress model and a repulsive force model.

We first introduce the viscous stress model.
The viscous shear stress $\tau_w$ at the wall is $\mu\partial{u}/\partial{r}$, where $\mu$ is the dynamic viscosity of the liquid, $u$ is the liquid velocity component in the direction parallel to the wall, and $r$ is the normal position to the wall.
In the present case, the velocity scale and the length scale for $\tau_w$ are $v$ and $D$, respectively.
%, resulting in Stokes drag during the jet penetration. %(\textcolor{red}{\bf CITATION?})
%The drag coefficient $C_D$ in this flow is known to be inversely proportional to Reynolds number, e.g. $C_D$=24/$Re$ for a rigid sphere in a Stokes flow. 
%Here $Re=vd/\nu$, where $d$ is the length scale of a object and $\nu$ is the kinematic viscosity of a liquid.
%The total viscous drag is the integral of the shear stress over total area.
In the viscous regime the relationship between the velocity and the drag force per unit mass is:
\begin{equation}
F_D = -c_{v}\cdot v,
\label{eq:F_D}
\end{equation}
where $c_{v}$ is a fitting parameter with the units of inverse time. %, including viscosity, geometrical effect, density effect, etc..
It is known that gentle deposition (small impact velocity)  into the soft matter gives no penetration.
The penetration starts when the impact velocity $v_{jet}$ exceeds a critical velocity $v_c$.
For $v_{jet} \ge v_c$ the final penetration depth $D_p$ is given by:
\begin{equation}
D_p = \frac{1}{c_{v}}(v_{jet}-v_c).
\label{eq:D_p}
\end{equation}
%where $v_{jet}$ is the initial velocity and $v_c$ is the critical velocity at which the penetration starts.

%\textcolor{blue}{
The other model considers a repulsive force acting on the jet.
%Due to the ultra-high velocity of the jet, the Reynolds number $Re$ for this jet can reach in the order of 10$^6$.
%The drag coefficient $C_D$ of an object in the high $Re$ flow is almost constant (e.g. $C_D$ $\sim$ 0.5 for a rigid sphere and $C_D$ $\sim$ 0.8 for a long cylinder in high $Re$).
The repulsive force is modeled as being proportional to the inertial force of the jet ($\sim \rho v^2$).
%For simplicity we only consider the first order of this force.
In this inertial regime, the relationship between the velocity and the drag force per unit mass is given by:
\begin{equation}
F_D = -c_{i}\cdot v^2,
\label{eq:Repulsive}
\end{equation}
where $c_i$ is a fitting parameter with dimensions of inverse length. % for this model.
Including the offset due to the critical velocity, we obtain the final penetration depth $D_p$ as:
\begin{equation}
D_p = \frac{1}{c_{i}}\ln (\frac{v_{jet}}{v_c}-1).
\label{eq:D_p_repulsive}
\end{equation}
%}

Both models are compared with the experimental results in figure~\ref{fig:models}.
It shows that the viscous stress model gives the best agreement, indicating that the jet likely experiences shear stress in the material.
This model gives predictive power to our novel method, enabling us to link the physical parameters of our lab experiments to real world medical applications.

\begin{figure}[h!]
\centerline{\includegraphics[width=0.45\textwidth]{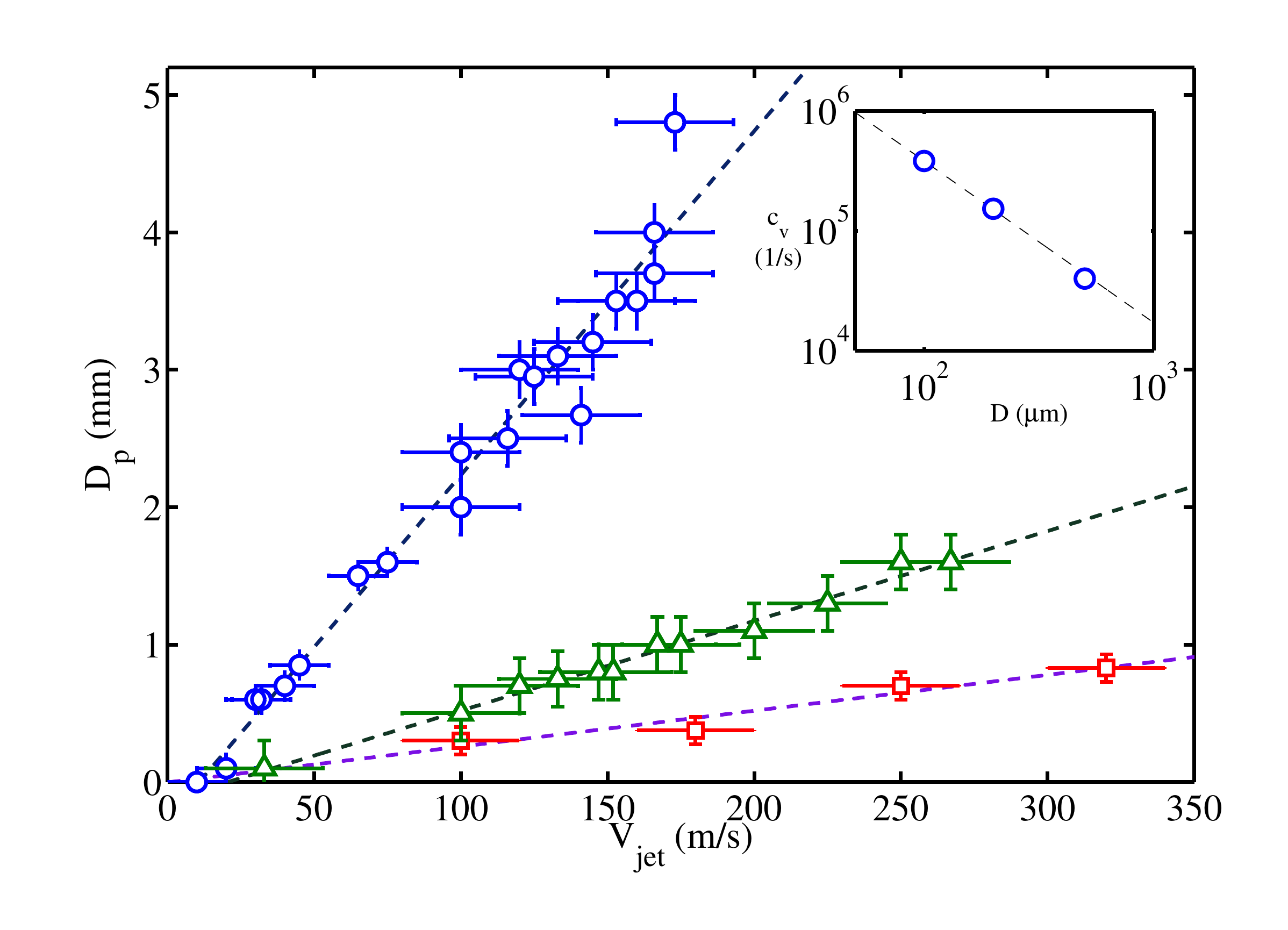}}
\caption{Injection depth as a function of the jet velocity for different capillary tube diameters. The data for each capillary tube are fitted by the viscous stress model, shown by the dashed line. The fitted slope $c_v$ is plotted as a function of the capillary size in the inset. 
The data can be represented by $c_v \propto D^{-1.35\pm 0.48} $, shown as the black dash-dotted line in the inset. \label{fig:TimeEvo}}
\end{figure}

Figure~\ref{fig:TimeEvo} shows the penetration depth for the jets created in tubes with three different diameters.
As discussed in \cite{Tagawa2012}, tubes with larger diameters result in jets with larger diameters.
At the same time, the penetration depth increases with the diameter of the jet.
For the 500 $\mu$m tube case, the jet can penetrate up to $\sim$ 5 mm in a single shot, something that has never been achieved so far.
For the 100 $\mu$m tube case, the jet penetrates 0.5 mm at a velocity of 320 m/s, close to sonic speed.
Thanks to the large velocity range of our jets, we can achieve the same penetration depth by using different tube diameters, which enables us to control the injection volume with nano-liter precision.

\begin{figure}[h!]
\centerline{\includegraphics[width=0.45\textwidth]{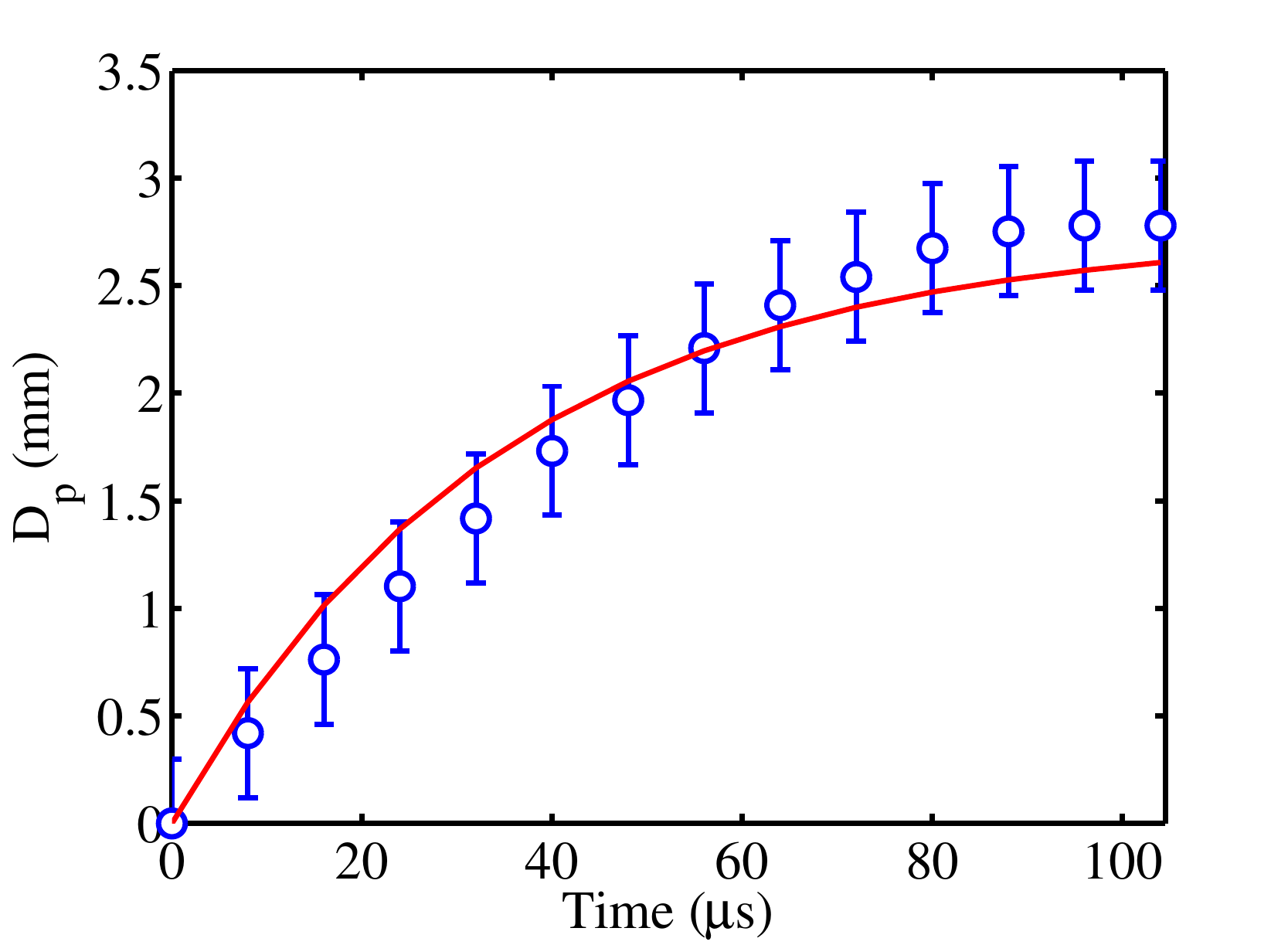}}
\caption{
The time evolution of the penetration depth $D_p$. % as a function of time. 
The blue markers show the experimental results for the jet of $v_{jet}$ = 120 m/s for the 500 $\mu$m tube.
The red thick line shows the penetration depth from equation~(\ref{eq:D_p(t)}) with $c_v$ obtained by fitting the model by equation~(\ref{eq:D_p}) to all data for the 500 $\mu$m tube shown in figure~\ref{fig:TimeEvo}.
\label{fig:DepthTime}}
\end{figure}

All data sets show a linear relation between the penetration depth and the jet velocity.
Remarkably the viscous stress model (equation~\ref{eq:D_p}) discussed above holds for all cases.
%for each data set is presented in the figure~\ref{gelatin}.
We will now try to calculate the parameter $c_v$ in equation~(\ref{eq:D_p}) from the jet geometry:
We approximate the jet shape by a cylinder, whose mass $m_j = \pi\rho D^2 l/4$, where $l$ is the length of the jet cylinder.
The total viscous stress on the jet is $F_v = \int_{dA} \tau dA$, where $A$ is the area on which the viscous shear stress acts. %, where $\tau_0 = \mu_g \partial{u}/\partial{y}|_{y=0}$ ($\mu_g$ is the dynamic viscosity of the gelatin in the direction $y$ which is proportional to the stream direction of the jet. At the jet-gelatin interface $y$ = 0.).
The elongated shape of the jet, i.e. the aspect ratio $D/l \ll$ 1 allows us to approximate $A\approx \pi D l$, leading to $F_v \sim \mu v l$.
Thus we obtain equation~(\ref{eq:F_D}) $F_D \sim c_v\cdot v$ with $c_v \propto D^{-2}$, meaning that $c_v$ quadratically decreases with increasing $D$.
% force balance and the shear inside the jet, $F_D \sim  \mu (u-u_s)/\rho D^2 \sim c_v\cdot u$.
%This relation and the fitted value $c_v $ allow us to assume the order of the slip velocity $u_s \sim O(10^{-3})\cdot u$.
%This leads $\mu_g \sim O(10^2)$, which is very high viscosity as a fluid, e.g. $\sim$ 100 times more viscous than honey.
%This $\mu_g$ also provides $Re = O (10^{-2})$, being consistent with our Stokes flow approximation.
%The corresponding fitting parameter $c_v$ is expected to be inversely proportional to jet diameter as discussed above.
The experimental values $c_v$ for each tube jet are shown in the inset in the figure~\ref{fig:TimeEvo}.
Indeed, larger values for $c_v$ are found for smaller $D$.
Assuming a power low $c_v\propto D^{\alpha}$ we obtain from the experimental results $\alpha$ = -1.35 $\pm 0.48$, slightly larger than the model result $\alpha$ = -2. %, having a visible discrepancy from predicted slope of -2 due to over-simplicity of the argument.

We also experimentally measure the time evolution of the penetration depth  and compare it with the viscous stress model. The model leads to an exponential temporal evolution of the penetration depth $D_p(t)$,
\begin{equation}
D_p(t) = \frac{v_{jet}-v_c}{c_{v}}(1-e^{-c_{v}t}).
\label{eq:D_p(t)}
\end{equation}
%where $v_{jet}$ is the initial velocity.
The viscous stress model shows an agreement with the measurement within error bars as shown in figure~\ref{fig:DepthTime}.
%The small deviation in the early stage could be due to the gelatin deformation which is not considered in the model.
This result provides additional support in favor of the viscous stress model for gelatin.\\
%To validate this viscous stress model further, we measure the depth of the hole as a function of time and compare these values with the equation~\ref{eq:D_p(t)}.
%Again, good agreement is found in figure~\ref{eq:D_p(t)}.

\subsection{Injection into artificially grown human skin}

\begin{figure}
\centerline{\includegraphics[width=.5\textwidth]{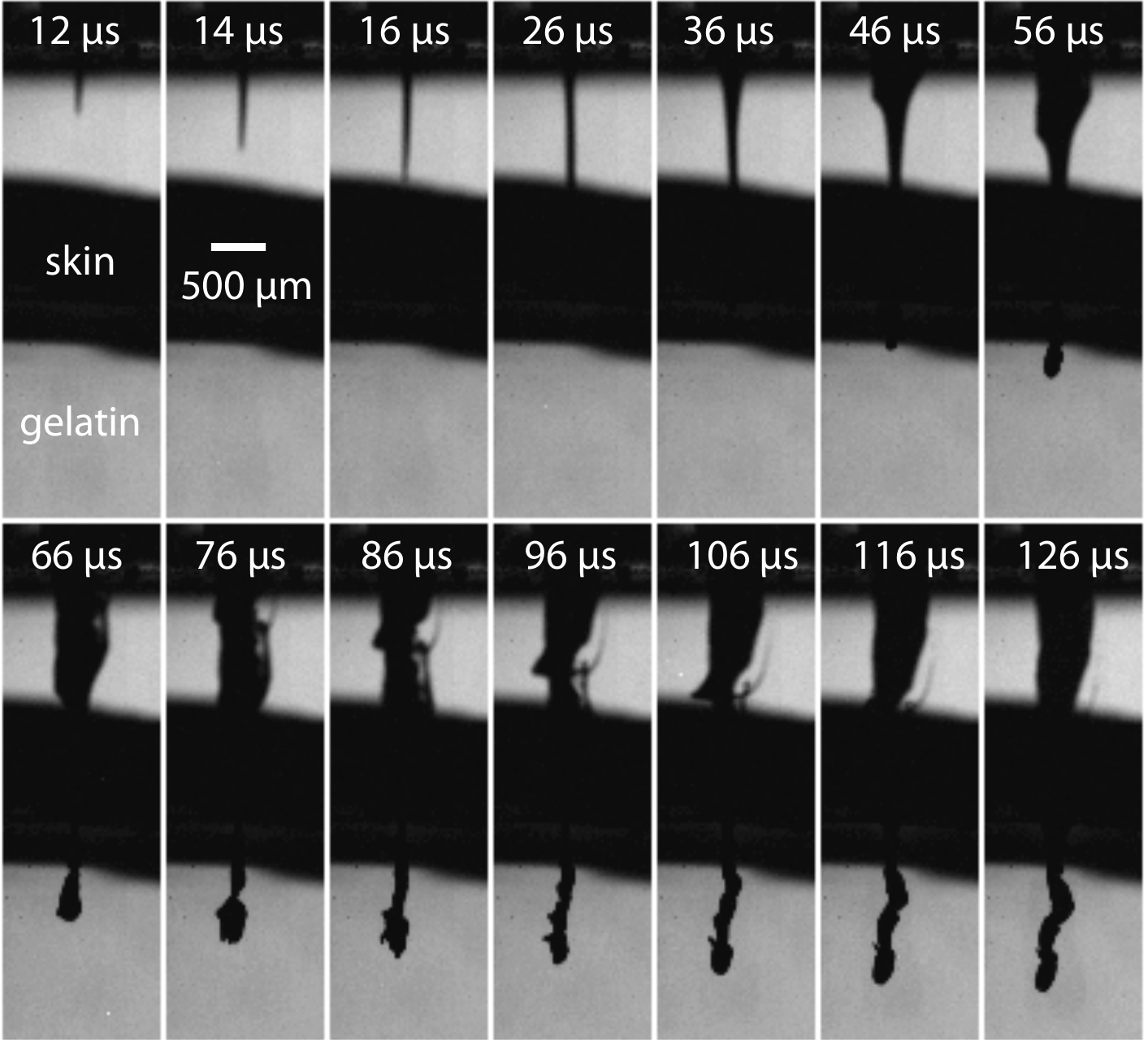}}
\caption{Time evolution of the jet penetration into human skin placed on gelatin. After the second snapshot, the time interval for each image is 10 $\mu$s. The laser is shot at 0 $\mu$s. The jet impact velocity is 160 m/s. Note that the dark region of the skin in the images is thicker than the actual thickness of the skin, as the skin curls up on the sides of the cuvette.
\label{ArtificialSkin}}
\end{figure}

In order to mimic the real human body, we have used artificially grown human skin placed on top of the gelatin 5 wt\% as a target material for our jets.
Figure~\ref{ArtificialSkin} shows snapshots of the jet penetrating into this material comprising of both the skin and the gelatin.
%From side view, it is not possible to observe the penetration dynamics in the skin.
The tip of the jet is observed for the first time in the gelatin at 46 $\mu$s.
At this point, it is clear that the jet is able to penetrate human skin.
After this stage, the penetration dynamics are similar to those in the gelatin case shown in figure~\ref{SAX}. The jet is still focused even though the jet has to penetrate through the additional barrier of skin.

\begin{figure*}[ht]
\centerline{\includegraphics[width=1\textwidth]{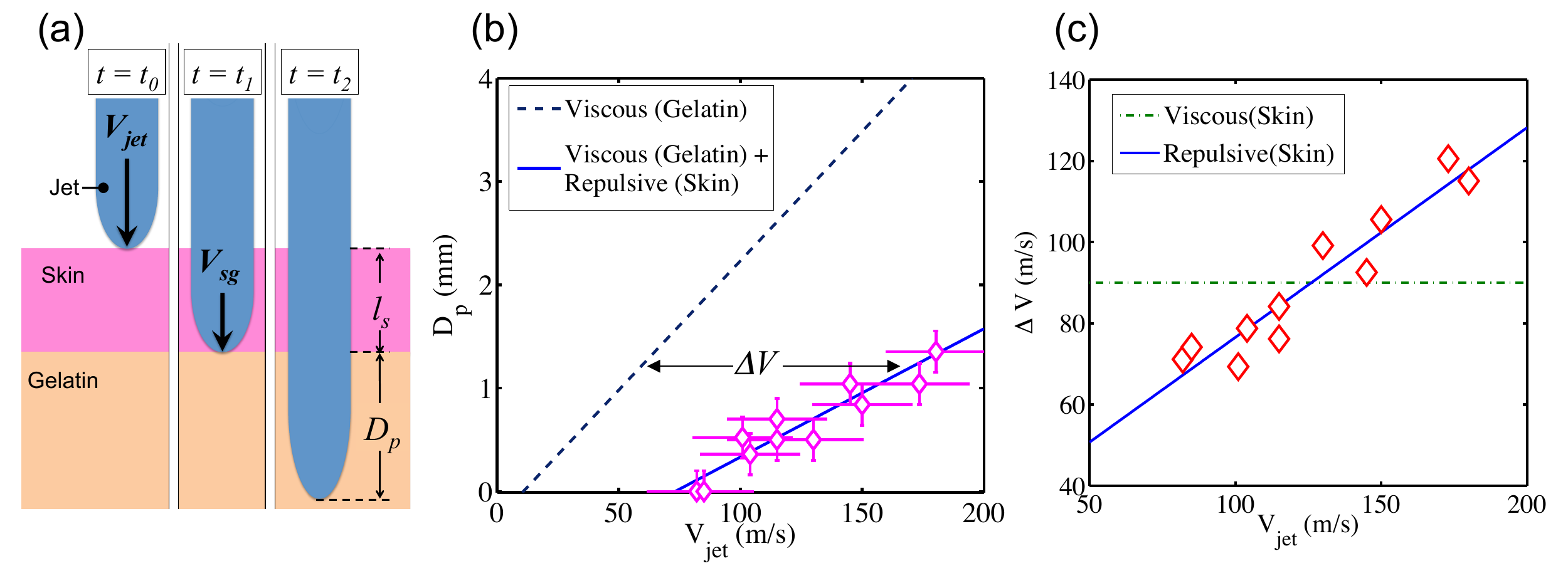}}
\caption{
(a) A sketch for the jet penetration into the gelatin covered with the artificially grown human skin layer (the thickness $l_s$ = 700 $\mu$m). 
At $t = t_0$ the jet impacts the skin layer with the velocity $V_{jet}$.
At $t = t_1$ the jet goes through the skin layer and start penetrating the gelatin with the velocity $V_{sg}$.
At $t = t_2$ the jet stops at the final depth $D_p$.
(b) Diamonds: the final depth $D_p$ into gelatin with the skin layer as a function of the jet velocity. % and is penetrated by the jet before entering into the gelatin. 
The dashed line: the case for gelatin without skin attached.
The blue thick line: the results of the model represented by equation~(\ref{eq:Iskin}).
(c) The velocity reduction $\Delta V$ due to the skin layer vs. the impact velocity.
The dark green line: the viscous stress model for the skin, and the blue thick line: the repulsive force model (equation~(\ref{eq:RepulsiveSkin})).
\label{D_pSkin}}
\end{figure*}

Figure~\ref{D_pSkin}(b) shows the penetration depth ($D_p$) into the gelatin through the skin as a function of the initial impact velocity ($V_{jet}$) of the jet.
Note that the data only represent the penetration depth ($D_p$) into the gelatin, excluding the skin thickness ($l_s$) as indicated in figure~\ref{D_pSkin}(a).
The threshold velocity for penetrating through the skin is found to be 80 m/s. %, indicating that the artificial skin is completely penetrated by the jet.
Even after the jet has penetrated an additional barrier in the form of human skin, the depth depends linearly on the initial velocity ($V_{jet}$). %jet velocity show a linear relation with the penetration depth.
This suggests an excellent controllability of this system, which is crucial for medical applications. 
As seen in figure~\ref{D_pSkin}(b), the jet can penetrate more than a millimeter into the soft tissues (gelatin in the present case), after passing through the skin barrier. 
This depth is sufficient for most medical applications, e.g. insulin injection or vaccinations.
When compared to the pure gelatin case (dashed line in figure~\ref{D_pSkin}(b)), the penetration depth is of course smaller with the skin layer being present. % slope of the fitted line is smaller.
%
%It indicates that the skin induces more drag force to the jet than that resulting from pure gelatin.
%Based on the viscous drag model of the pure gelatin for the 500 $\mu$m tube, we can calculate the jet velocity just before impacting on the gelatin $v_{gel}$.
%The estimated $v_{gel}$ as a function of the initial velocity $v_{jet}$ is shown in the inset of figure~\ref{D_pSkin}.
%The $v_{gel}$ increases with increasing initial velocity $v_{jet}$.
The skin decelerates the jet until complete penetration through itself (at $t = t_1$ as shown in figure~\ref{D_pSkin}(a)), after which the jet penetrates the gelatin until complete stoppage (at $t = t_2$).
For the latter process we again adopt the viscous drag model, but with a reduced velocity $v_{sg}$ due to the additional barrier of the skin (as shown in figure~\ref{D_pSkin}(a)).
The velocity reduction by the skin layer is 
\begin{equation}
\Delta V = V_{jet}-V_{sg},
\end{equation} 
which, as indicated in figure~\ref{D_pSkin}(b), is equivalent to the horizontal offset between the line for the pure gelatin and the measured data with the skin layer being present.
Figure~\ref{D_pSkin}(c) shows this velocity reduction ($\Delta V$) as a function of the impact velocity ($V_{jet}$).\\ We evaluate the velocity reduction by considering the drag of the skin layer again with two different models: the viscous stress model and the repulsive force model.
%now consider the two drag models again for the skin layer: the viscous stress (equation~(\ref{eq:F_D}))  and the repulsive force (equation~(\ref{eq:Repulsive})).
The viscous stress model for the skin layer leads to a constant velocity reduction for a given skin thickness (see equation~\ref{eq:D_p}).
However, the experimental results show a different trend in figure~\ref{D_pSkin}(c).
Hence, we model the skin layer with a repulsive force (see equation~\ref{eq:Repulsive}), and the corresponding velocity reduction is
\begin{align}
\Delta V= v_{jet}-(v_{jet} - v_s)e^{-c_{i,s}l_s},
\label{eq:RepulsiveSkin}
\end{align}
with two fitting parameters: $c_{i,s}$ and $v_s$, and the skin thickness $l_s$ = 700 $\mu$m.
%Both models are fitted to the $\Delta V$ calculated from experiments as shown in figure~\ref{D_pSkin}(c).
Figure~\ref{D_pSkin}(c) reveals that the model described by equation~(\ref{eq:RepulsiveSkin}) with $c_{i,s}$ = 1.0$\cdot10^{3}$ m$^{-1}$ and $v_s$ = 51.5 m/s shows a good agreement with the experiments, suggesting that the jet experiences the repulsive force in the skin layer.
This is probably due to the increased hardness of the skin compared to that of gelatin.
%Here $v_s$ is the critical velocity for penetration through the skin, $c_{v,s}$ and  $c_{i,s}$ are fitting parameters, and the skin thickness $l_s$ is 700 $\mu$m.
%After penetrating through the skin layer, the jet starts penetrating the gelatin, which is very similar to the pure gelatin case. 
The final penetration depth inside the gelatin can be therefore obtained as 
% by substituting these values of $v_{sg}$ into equation~(\ref{eq:D_p}). 
%Using the viscous drag model for the skin layer we obtain,
%\begin{subequations}
%\begin{align}
%D_{p} = \frac{1}{c_{v}}(  v_{jet} - v_s - c_{v,s}D_s   - v_c),
%\label{eq:Vskin}
%\end{align}
%and the repulsive force model for the skin layer yields,
\begin{equation}
D_{p} = \frac{1}{c_{v}}( (v_{jet} - v_s)e^{-c_{i,s}l_s} - v_c),
\label{eq:Iskin}
\end{equation}
which is plotted as the thick line in figure~\ref{D_pSkin}(b).
%\end{subequations}
%Both models are fitted to the measurements as shown in figure~\ref{D_pSkin}.
The excellent agreement suggests that the model (equation~(\ref{eq:Iskin})) combining the repulsive force for the skin layer and the viscous drag for the gelatin, nicely describes the depth of the jet penetration.  
This model is thus suited to quantitatively describe the penetration of high-speed jets into human skin enclosing soft tissue.

%
%with the experimental data. % compared to that by equation~(\ref{eq:Vskin}).
%
%\textcolor{blue}{
%Combining the models for the skin and the gelatin equivalent to the human soft tissue, the penetration depth in this case becomes
%As shown in  figure~\ref{D_pSkin} this model well describes the depth of the penetration into human skin on the soft tissue.
%}

In this study we have shown that a novel method for needle-free injections can resolve many of the longstanding issues that have prevented largescale adaptation of needle-free injection systems. 
We show that a highly-focused geometry of the jets and a wide range of velocities is essential for good controllability, versatility, and effectiveness of needle-free injection systems.
We also model the penetration of the jet into soft matter and human skin enclosing soft tissue. 
The results presented here take needle-free injections a step closer to widespread use.

\begin{appendix}
\section*{Materials}
%\section{Materials and methods} \\

\subsection{Microjet Generation}
The highly focused high-speed microjets are generated by focusing a laser pulse into a small capillary tube filled with water-based red dye. 
This leads to the abrupt vaporization of a small mass of liquid \cite{Sun2009}. 
The vaporization causes a shock wave to travel through the liquid and impulsively accelerate the curved liquid interface due to kinematic focusing. 
The capillary tube is connected to a syringe through micro tubing and the dye is pumped into the capillary tube using a syringe pump. 
The characteristics of this jet, such as velocity and width, can be controlled by varying the laser power, the distance between the laser focus and the free surface, the liquid-glass contact angle, and the diameter of the tube \cite{Tagawa2012}. \\
\subsection{Injection Into Gelatin}
Gelatin mixtures were used to study the injection into solid substrates. 
The gelatin was prepared a few hours before the experiments by dissolving 5 weight \% of gelatin in MilliQ water. 
After dissolving the gelatin, the mixture was poured into small 1 cm$\times$1 cm cuvettes and put in the fridge (4\,$^{\circ}$C) for an hour. Penetration dynamics were filmed using high-speed cameras with frame rates up to 10$^6$ fps (HPV-1, Shimadzu Corporation, Japan, and FASTCAM SAX, Photron, USA). \\
\subsection{Penetration Across Human Skin In Vitro}
The artificial skin was cultured by the Department of Dermatology of the Leiden University Medical Center. The Leiden Human Epidermal (LHE) skin model used in this study has been fully characterized and shows very high similarities with native skin \cite{Ghalbzouri2008}. 
The LHE represents a full-thickness model (epidermis generated onto a dermal matrix).
The skin was supplied in patches of 2.4 cm in diameter and was kept in an incubator prior to experiments. For the penetration experiments, the skin layers were placed on top of the gelatin mixtures in the small cuvettes. 
Penetration dynamics were filmed using high-speed cameras (HPV-1, Shimadzu Corporation, Japan, and FASTCAM SAX, Photron, USA). \\
\subsection{Velocity and Depth Measurement}
High-speed cameras (HPV-1, Shimadzu Corporation, Japan, and FASTCAM SAX, Photron, USA) were used to record the injection process. 
The velocity and depth were determined from these high-speed recordings.

%\begin{figure}[h!]
%	\centerline{\includegraphics[width=0.3\textwidth]{penetration}}
%	\caption{Experimental setup of injection trials. The highly focused high-speed is shot towards the reference materials. The gelatin was prepared by dissolving 5 weight \% of gelatin in MilliQ water, which simulates the properties of soft tissue in the human body \cite{Menezes2009}. The artificial skin (The Leiden Human Epidermal (LHE) skin model), showing very high similarities with native skin \cite{Ghalbzouri2008}, is placed on the gelatin to resemble real human body.   
%}
%	\label{setup}
%\end{figure}

%\end{materials}
\end{appendix}

\begin{acknowledgments}
We thank C. Clanet,  F. Dijksman, B. Hoeksma, L. Homan, Devaraj van der Meer, Vivek N. Prakash, and C.W. Visser for fruitful discussions.
We appreciate the financial support given by Fundamenteel Onderzoek der Materie (FOM), which is part of Nederlandse Organisatie voor  Wetenschappelijk Onderzoek (NWO) and the European Research Council (ERC) through a Proof of Concept Grant.

\end{acknowledgments}

%\end{article}

\end{document}